\documentclass[aps,twocolumn,prb]{revtex4}

\newcommand{\ket}[1]{\left |#1\right >}

\newcommand{\expec}[1]{\left < #1\right >}

\usepackage{amsmath}
\usepackage{amsfonts}
\usepackage{graphicx}
\usepackage{amsthm}
\usepackage{subfigure}
\usepackage{hyperref}
\usepackage{ulem}

\usepackage[usenames]{color}

\renewcommand{\v}[1]{\boldsymbol{#1}}

\newcommand{\slashuparrow}{\!-\!\!\!\!{\uparrow}}
\newcommand{\slashdownarrow}{\!-\!\!\!\!{\downarrow}}
\newcommand{\lp}{\left ( }
\newcommand{\rp}{\right ) }

\newcommand{\beq}{\begin{eqnarray*}}
\newcommand{\eeq}{\end{eqnarray*}}

\newcommand{\be}{\begin{eqnarray}}
\newcommand{\ee}{\end{eqnarray}}

\def\lsim{\mathrel{\rlap{\lower4pt\hbox{\hskip1pt$\sim$}}
    \raise1pt\hbox{$<$}}}                % less than or approx.

\def\gsim{\mathrel{\rlap{\lower4pt\hbox{\hskip1pt$\sim$}}
    \raise1pt\hbox{$>$}}}                % greater than or approx.

\begin{document}

\title{Candidate theories to explain the anomalous spectroscopic signatures of atomic  H  in molecular  H$_2$ crystals}
\author{Kaden R.A. Hazzard}
\email{kh279@cornell.edu}
\affiliation{Laboratory of Atomic
and Solid State Physics, Cornell University, Ithaca, New York 14853}
\author{Erich J. Mueller}
\affiliation{Laboratory of Atomic
and Solid State Physics, Cornell University, Ithaca, New York 14853}

\begin{abstract}
We analyze a number of proposed explanations for spectroscopic anomalies observed in atomic hydrogen defects
embedded in a solid molecular hydrogen matrix.  In particular, we critically evaluate the possibility that these anomalies are related to Bose-Einstein condensation (both global and local).  For each proposed mechanism we discuss which aspects of the experiment can be explained, and make predictions for future experiments.
\end{abstract}

\maketitle

\section{Introduction and motivation}

Quantum solids, where the zero point motion of the atoms is greater than roughly 10\% of their separation, form a fascinating class of materials.
A principal question with these materials is to what extent they are quantum coherent, and under what conditions they can be supersolid -- supporting dissipationless mass flow.
Examples of quantum solids include
$^4$He~\cite{prokofiev:supersolid-review,kim:prob-obs-of-supersolid,kim:supersolid-obs-of-superflow,rittner:disorder-confinement,rittner:annealing-supersolid}, solid hydrogen~\cite{kumada:tunnel}, Wigner crystals, and atomic hydrogen defects in solid molecular hydrogen~\cite{ahokas:solid-hydrogen-anomalies-PRL,vasilyev:solid-hydrogen-anomalies-lt24-proc,nature-overview:solid-hydrogen}.
Here we theoretically study the last system, giving a critical evaluation of scenarios of Bose-Einstein condensation of atomic hydrogen defects.  We make testable predictions for these scenarios.

Solid hydrogen, with a Lindemann ratio of 0.18,  is the only observed molecular quantum crystal.
It is a rich system with
many rotational order/disorder transitions~\cite{silvera:review}.  %p. 429
The phenomenology of H$_2$ solids is even more interesting when atomic H defects are introduced~\cite{ahokas:solid-hydrogen-anomalies-PRL,vasilyev:solid-hydrogen-anomalies-lt24-proc,nature-overview:solid-hydrogen}.  Here we focus on the spectroscopic properties of this system, and how they may be related to quantum coherence.

Our work is motivated by recent experiments in low-temperature ($T\sim150$mK) solid molecular hydrogen, populated with large densities ($n\sim  10^{18} \text{cm}^{-3} $) of atomic hydrogen defects.  These experiments observe unexplained internal state populations~\cite{ahokas:solid-hydrogen-anomalies-PRL,vasilyev:solid-hydrogen-anomalies-lt24-proc,nature-overview:solid-hydrogen}.
Ahokas \textit{et al.}~\cite{ahokas:solid-hydrogen-anomalies-PRL} provocatively conjectured that the anomalies may be related to Bose-Einstein Condensation (BEC) of the atomic defects.  Here our goal is to explore and constrain this and alternative scenarios.  We conclude that global Bose-Einstein condensation could not realistically explain the experiments, and that local condensation would lead to distinct signatures in future experiments.

\section{Experiments}

We review Ahokas \textit{et al.}'s~\cite{ahokas:solid-hydrogen-anomalies-PRL} experimental apparatus, results, and observed anomalies.

\subsection{General introduction: physics of atomic hydrogen embedded in solid hydrogen}

\begin{figure}[hbtp]
\setlength{\unitlength}{1.0in}
\centering
\includegraphics[width=3.2in,angle=0]{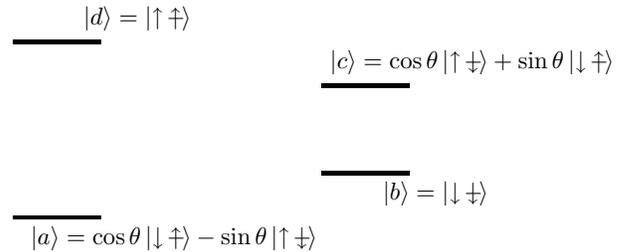}
\caption{  Hyperfine level diagram for hydrogen atom in a $B=4.6$T magnetic field, where the mixing angle is $\theta=3 \times 10^{-3}$.
Vertical axis schematically denotes energy, while horizontal axis has no physical meaning.
Arrows denote electron (no-slash arrow) and nuclear (slashed arrow) spin projections.
\label{fig:hyperfine-h}
}
\end{figure}

\textbf{Hyperfine structure of atomic hydrogen.}
Fig.~\ref{fig:hyperfine-h} schematically shows the level structure for a H atom in a $B=4.6$T magnetic field, similar to that used in the experiments of interest~\cite{ahokas:solid-hydrogen-anomalies-PRL}.  At these large fields, the levels break into two nearly degenerate pairs. Levels within a pair are separated by radio frequencies and the pairs are separated by microwave frequencies.
The electronic spin in states $a$ and $b$ is aligned with the magnetic field and is anti-aligned in the other states.

Ahokas \textit{et al.}~\cite{ahokas:solid-hydrogen-anomalies-PRL} observe that within the solid hydrogen matrix the atomic hydrogen's spectra is modified. The a-d energy splitting decreases while the b-c energy splitting increases by the same amount.

\textbf{Spectroscopy of atomic hydrogen.}
We consider three different spectroscopic probes, distinguished by the states they connect: nuclear magnetic resonance (NMR), electron spin resonance (ESR), and electron-nuclear double resonance (ENDOR).  NMR uses radio waves to couple the $a$ and $b$ states, while ESR uses microwaves to couple the $a$ and $d$ or $b$ and $c$ states.  ENDOR uses a two photon transition to couple  $a$ and $c$.  All of these probes can be used in the linear regime, where they give information about the splittings and occupation numbers, or in the non-linear regime.  As an example of the latter,  Ahokas \textit{et al.}~\cite{ahokas:solid-hydrogen-anomalies-PRL} study what happens when they apply high RF power, saturating the NMR line.

\textbf{Molecular hydrogen.}
At these temperatures, two states of molecular hydrogen are relevant: the ``para" and ``ortho" states.
In the ``para" or ``p-" configuration the relative nuclear wavefunction $\psi$ is symmetric under exchanging nuclei and the nuclear spins form a singlet. In the ``ortho" or ``o-" configuration  $\psi$ is antisymmetric and the nuclear spins form a triplet.
In all cases, the electrons are in a symmetric bonding orbital and the electronic spins are consequently anti-aligned.

The true ground state of solid hydrogen is formed of p-H$_2$. In practice, however, it is very rare to have a pure p-H$_2$ sample.  The ortho state is long lived\cite{silvera:review}, requiring hours for the concentration to change by  1\%.  This ortho-para conversion can be a source of heating in experiments with an energy  $\Delta/k_B = 170$ K released per molecule.  In the experiments of Ahokas \textit{et al.} \cite{ahokas:solid-hydrogen-anomalies-PRL} the exact quantity of o-H$_2$ is unknown, but given the growth technique it is likely to be at least ten percent.

At standard pressure the p-H$_2$ in the solid is highly spherical:  interaction with neighboring H$_2$ negligibly distorts the p-H$_2$.  Modeling the hydrogen-hydrogen interactions by their vacuum values quite accurately describes quantities such as the speed of sound~\cite{silvera:review}.
If sufficient ortho-hydrogen is present, orientational ordering transitions may occur around 1K.  The models we consider do not rely upon any orientational ordering.
Depending on sample-preparation conditions, either hcp or fcc crystals may be produced~\cite{silvera:review}.

\textbf{Atomic hydrogen in the solid lattice.}
The motion of atomic hydrogen in molecular hydrogen has been widely studied~\cite{kumada:tunnel}.  Both thermally activated and quantum tunneling contribute to defect motion, but quantum tunneling dominates at these low temperatures (the two dominant tunneling pathways have energy barriers of 4600K and 100K).   One motivation for these studies is to explore the possibility of Bose-Einstein condensation of these defects.  This is conceptually related to supersolidity driven by condensation of vacancies.

Kumada~\cite{kumada:tunnel} argues on the basis of experimental data that the exchange reaction $H+H_2 \rightarrow H_2 + H$ is the dominant diffusion mechanism at low temperatures. Other tunneling  pathways are possible, including  correlated, collective relaxation  and ``physical" diffusion.

Ahokas \textit{et al.}~\cite{ahokas:solid-hydrogen-anomalies-PRL} achieve populations of 50ppm H defects in their solid, and argue that H sits at substitutional sites.  The observed lifetime of these defects was weeks.  The dominant decay mechanism should be the recombination of two hydrogen defects. One therefore  expects that this rate is determined by the diffusion rate of the defects. The long lifetime is therefore inconsistent with the diffusion rates predicted by phonon assisted tunneling.  One possible explanation is the suppression of tunneling by the strain-induced mismatch of energy levels between neighboring sites~\cite{kagan:slow-tunnel-strain-mismatch,ivliev:slow-tunnel-strain-mismatch}.

\textbf{Crystal growth.}
Ahokas \textit{et al.}~\cite{ahokas:solid-hydrogen-anomalies-PRL} grow solid hydrogen from a gas of electron-spin polarized metastable hydrogen atoms, which undergo two-hydrogen recombination to form molecules -- these events are only allowed in the presence of walls.  The  H$_2$ solid grows layer-by-layer from at a rate of 0.5-1 molecular layer per hour.  After $\sim1$week, a quartz microbalance revealed a  film thickness of $150\pm1$ layers, \footnote{private communication with Jarno J\"arvinen}.

\subsection{Anomalies and experimental results\label{sec:anomalies}}

 Ahokas \textit{et al.}~\cite{ahokas:solid-hydrogen-anomalies-PRL} observed four anomalies:
(1) several orders-of-magnitude too fast ``Overhauser" relaxation,
(2) a non-Boltzmann a-b population ratio,
(3) saturation of the a-b spectroscopic line fails to give a  1:1 population ratio, and
(4) recombination rates are extremely low.
Items 2 and 3 will be  our main focus.

\textbf{Overhauser relaxation.}
The $c$ to $a$ relaxation is expected to be extremely small at the $4.6$T fields of the experiments.  This can be seen from the small mixing angle $\theta \approx 3 \times 10^{-3}$,  indicated in Fig.~\ref{fig:hyperfine-h}.  The mixing angle appears in the states as
$\ket{a} = \cos \theta \ket{\downarrow \slashuparrow} - \sin \theta \ket{\uparrow \slashdownarrow},$ and
$\ket{b} = \cos \theta \ket{\uparrow \slashdownarrow} - \sin \theta \ket{\downarrow \slashuparrow}$.
Any c-a decay mechanism by photon emission is suppressed by $\theta^\alpha$ with~\cite{prokofev:overhauser-rates}  $\alpha \gsim 1$.  Ahokas \textit{et al.}~\cite{ahokas:solid-hydrogen-anomalies-PRL} observe no such suppression: the c-a line decays with a time constant of $\lsim 5$s, similar to the d-a decay time.

\textbf{Equilibrium populations.}
The polarization
\be
p &\equiv& \frac{n_a-n_b}{n_a + n_b}\label{eq:polarization-defn}
\ee
characterizes the a and b state population.
Assuming a Boltzmann distribution
$n_a/n_b = \exp\lp\Delta_{ab}/T\rp$ where $\Delta_{ab}\approx 43$mK is the difference in energies between the b and a states, one expects
%$ p = \tanh \lp \frac{\Delta_{ab}}{T}\rp %\label{eq:polarization-temp-dep} $
%so
$p=0.14$
at 150~mK.
On the contrary, Ahokas \textit{et al.}~\cite{ahokas:solid-hydrogen-anomalies-PRL} measure $p=0.5$.

When one of the states is depleted
%a strong rf or microwave field is used to drive the population away from this measured polarization,
the system returns to this non-Boltzmann value on a time scale of $\sim 50$hours.

\textbf{Saturation of a-b line.}
An extremely strong rf field should saturate the a:b line, driving the population ratio to 1:1, or $p=0$.  Ahokas \textit{et al.}~\cite{ahokas:solid-hydrogen-anomalies-PRL} obtained a minimum of $p=0.2$ at high excitation powers.  For sufficiently large power this saturated population ratio was independent of the applied rf power.

\textbf{Low recombination rates.}
As previously described, at 150mK Ahokas \textit{et al.}'s~\cite{ahokas:solid-hydrogen-anomalies-PRL} recombination rate
is much smaller than expected, negligible on a time scale of weeks.  In contrast, at $T=1$K their recombination rates
are consistent with previous studies~\cite{ivliev:recomb-rate}.

\textbf{Hole burning.}
Ahokas \textit{et al.} applied a magnetic field gradient and a rf field to saturate the a-b line in a millimeter sized region of the sample.
The spectral hole recovered in a time similar to that in the homogeneous case, indicating that the nuclear spin-relaxation is somewhat faster than spin migration.  This would seem to indicate that the atomic hydrogen defects are immobile on regions much larger than a millimeter.

\section{Scenarios\label{theory}}

In this section we evaluate some previously proposed scenarios for these phenomena and suggest a new one~\cite{ahokas:solid-hydrogen-anomalies-PRL}.  For each scenario, we present the idea, examine its consistency with Ahokas \textit{et al.}'s~\cite{ahokas:solid-hydrogen-anomalies-PRL}  experiments, consider possible microscopic mechanisms, and \textit{give testable predictions.}

\subsection{Bose statistics and Bose-Einstein condensation\label{sec:Bose}}

\subsubsection{Idea}
Ahokas \textit{et al.}~\cite{ahokas:solid-hydrogen-anomalies-PRL} suggested
Bose-Einstein condensation (BEC) as a possible mechanism to explain the departure from the Boltzmann distribution.  In a BEC the lowest energy mode becomes ``macroscopically occupied," leading to an excess of $a$-H.
%Hydrogen BEC would lead to excess  $\ket{a}$ state population, as observed.
%Since the H's are effectively bosons in a lattice, they should condense at some temperature.
Neglecting interactions, the BEC transition temperature $T_c$ for a homogeneous system of spinless particles with density $\rho$ and effective mass $m^*$ is
\be
T_c &=& \lp \frac{\rho}{\zeta(3/2)}\rp^{2/3} \frac{2\pi\hbar^2}{k_B m^*}.\label{eq:BEC-Tc}
\ee
 Here $\zeta$ is the Riemann zeta function; $\zeta(3/2)\approx 2.61$~\cite{pethick:p-s-book}.
Including the b states is straightforward and makes only small changes:
%when $\Delta_{ab}$ is large the formula is exact, while
for example if $\Delta_{ab}=0$ one would divide the density by a factor of 2.  Note that as $m^*$ becomes larger the transition temperature becomes smaller.

\subsubsection{Phenomena explainable}
This scenario can in principle explain the non-Boltzmann equilibrium ratio $n_b/n_a$.  It does not provide an explanation of the inability to saturate the $a$-$b$ line, the slow recombination, or the fast Overhauser cross-relaxation.

\subsubsection{Consistency with experiment}

\textbf{Transition temperature and densities.}
As reported in Ahokas \textit{et al.}~\cite{ahokas:solid-hydrogen-anomalies-PRL},
for density $\rho\sim 10^{18}\text{cm}^{-3}$ and effective mass $m^*$ similar to bare mass $m$, the ideal Bose gas transition temperature in Eq.~(\ref{eq:BEC-Tc}) is
 $T_c\sim30$mK, far below the experimental temperature.
To address this inconsistency Ahokas \textit{et al.}~\cite{ahokas:solid-hydrogen-anomalies-PRL} suggest that phase separation may concentrate the defects to locally higher densities.  For example,
if the defects phase separated so that their density was $\rho \sim 3 \times 10^{19}$cm$^{-3}$, then with $m^*=m$ the transition temperature would be $T_c=170$~mK, and one would reproduce the observed ratio of $n_a/n_b$ at $T=150$~mK.  Such a powerful concentrating mechanism would have additional consequences, such as increased recombination rates.

A key question is how large the effective mass is.
In the following subsection we use the experimental hole-burning data to constrain the effective mass, finding that it is sufficiently large to completely rule out simple Bose-Einstein condensation of defects at the experimental temperatures.

\textbf{Estimate of effective mass.}
We expect that the effective mass $m^*$ of the defects is much higher than that of the free atoms.  Here we bound the effective mass by considering  Ahokas \textit{et al.}'s measurement of the lifetime of a localized spectral hole.  They found that a $w \sim 0.2$mm hole persisted for $\tau_{\text{pers}}>50$hours.  Our argument will relate macroscopic motion (which fills in the spectral hole) to microscopic motion (the defect tunneling).  We assume diffusive motion, where the characteristic time between collisions is longer than a tunneling time.

There are at least three mechanisms by which the spectral hole can heal: the excited atoms can spontaneously undergo a transition back to the $a$-state, spin exchange collisions can lead to spin diffusion, or $a$-state atoms can diffuse back into that region of space.
Neglecting all but the last process gives us an upper bound on the atomic diffusion constant $D$,
\be
D &\lsim& \frac{w^2}{\tau_{\text{pers}}} \approx  10^{-8} \text{cm}^2{/s}.\label{eq:diff-const-length-time}
\ee
Throughout this argument we will aim to produce an order-of-magnitude estimate, and use the symbol ``$\sim$" to indicate that we neglect constants of order unity.  This diffusion constant can be related to the microscopic collision time
$\tau_{\text{coll}}$ and the mean velocity $v$
by
\be
D &\sim& v \ell\label{eq:dc-vel-coll}.
\ee
with $\ell$ the mean free path.
%Moreover, we can bound $\tau_{\text{coll}}$ because in any physical situation, the collisions must occur on length scales longer than a lattice spacing $d$, so $\tau_{\text{coll}} \gsim d/v$.
In the effective mass approximation, one would expect thermal effects to yield a mean velocity
$v\sim\sqrt{k_B T/m^*}$.  An lower bound for the mean free path is given by the lattice constant
 $\ell\gsim d\sim 3$\AA~\cite{silvera:review}.  Thus we arrive at the following lower bound for the effective mass
 \be
m^* &\gsim& k_B T\lp \frac{ d}{10^{-8}\text{cm}^{2}/\text{s}}\rp^2 \approx 10^8 m_{\rm H},
\ee
where $m_H$ is the bare hydrogen mass and the last approximate equality is for the experimental temperature $T=150$mK.  This effective mass is several orders of magnitude too large to allow BEC at experimentally relevant temperature scales, regardless of the defect concentration.   This argument has neglected interactions and inhomogeneities:  phase separation or some ``local" BEC's that are uncoupled, thus disallowing global transport, would invalidate the arguments leading to our bound.

\subsubsection{Microscopic mechanism}

Microscopic estimates of tunneling matrix elements are beyond the scope of this work.  As previously discussed, the exchange reaction mechanism is expected to be the dominant pathway\cite{kumada:tunnel}.
%If they are as large as implied, then condensation is ruled out.   On the other hand, it is plausible that the tunneling matrix elements are much smaller, on the order of the bare mass.  This would imply that our simple picture of a hole recovery in a microscopically homogeneous system is incorrect.  More careful estimates of the tunneling are thus desirable to distinguish these scenarios.

\subsubsection{Experimental predictions}

\textbf{Polarization temperature dependence.} Perhaps the most easily testable prediction of this model is the temperature dependence of the polarization,
$p=(n_a/n_b-1)/(n_a/n_b+1)$ with
\be
\frac{n_a}{n_b} &=& \frac{\int d^3 k \,n[\epsilon_a(\v{k})/(k_BT)]}{\int d^3 k \,n[\epsilon_b(\v{k})/(k_BT)]}
\ee
with $n(x) \equiv 1/\lp e^x -1 \rp$.
The energy dispersion of atoms is
$\epsilon_a(\v{k})\approx\hbar^2k^2/2m^*$,
and
$\epsilon_b(\v{k})\approx\hbar^2k^2/2m^*+\Delta_{ab}$,
where $\Delta_{ab}$ is the b-a energy difference.
Above the BEC transition temperature $T_c$,
the integrals yield
\be
\frac{n_a}{n_b} &=& \frac{g_{3/2}\lp e^{\mu/T}\rp}{g_{3/2}\lp e^{\lp\mu-\Delta_{ab}\rp/T}\rp}\label{eq:pop-ratio-finite-mu}
\ee
where $g_\alpha(x)=\sum_j x^j/j^\alpha$ is the polylog function and $\mu$ is self-consistently determined to set $N$, for a homogeneous, three-dimensional gas.
The same expression holds in dimension $d$ with $3/2$ replaced by $d/2$.
Below
$T=T_c$,
one instead finds
\be
\frac{n_a}{n_b} &=& \lp\frac{T_c}{T}\rp^{3/2}\frac{\zeta(3/2)+g_{3/2}\lp e^{-\Delta_{ab}/T_c}\rp}{g_{3/2}\lp e^{-\Delta_{ab}/T}\rp}-1.\label{eq:pop-ratio-v-T}
\ee
While the effective mass sets the density at $T_c$, it does not appear in this expression.  The polarization depends only on $\Delta_{ab}/T$ and $T/T_c$.  To produce the observed $n_a/n_b=3$ at $T=150$mK, one needs $T_c=170$mK.

\begin{figure}[hbtp]
\setlength{\unitlength}{1.0in}
\centering
\includegraphics[width=3.2in,angle=0]{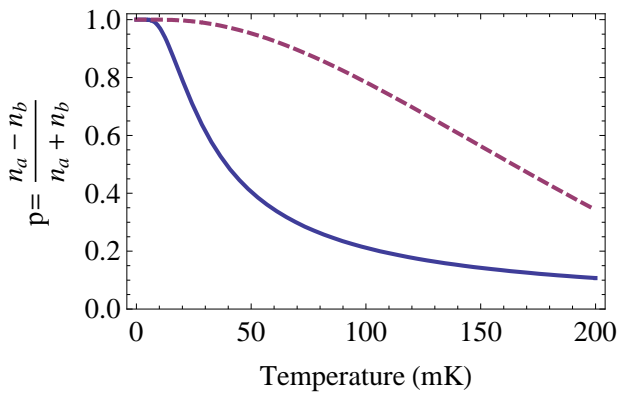}
\caption{ The polarization versus temperature for the Boltzmann case (solid line) and the Bose-condensed case (dashed line), from Eq.~\ref{eq:pop-ratio-v-T}. \label{fig:BEC-temp-dep}
}
\end{figure}

Fig.~\ref{fig:BEC-temp-dep} shows
$p(T)$ for Boltzmann and Bose condensed (assuming $T_c=200$mK) gases.  Accurately measuring $p(T)$ would clearly distinguish Bose and Boltzmann statistics.

Equation~(\ref{eq:pop-ratio-finite-mu}) shows that Bose statistics can affect the ratio $n_a/n_b$ even if the system is non-condensed.  However, Eq.~(\ref{eq:pop-ratio-finite-mu}) bounds $n_a/n_b<\zeta(3/2)/g_{3/2}(e^{-\Delta_{ab}/T})=2.3$, which is insufficient to explain the experimentally observed $n_a/n_b=3$.

\textbf{Transport.}
A second signature of BEC
is superflow.
For example, the sample could be incorporated into a torsional oscillator, providing a measurement of a possible nonclassical moment of inertia $I$:
 below $T_c$, the superflow decouples from the cell,
 and $I$ decreases.
 Mounting the sophisticated hydrogen growth and measurement equipment
 in an oscillator would be challenging, as would the difficulty of working with such small samples.

\textbf{Bimodal Cold Collision Shifts.}
  Bose Einstein condensation also has implications for the cold collision shifts in the atomic spectra.
  %  In particular, interactions shift spectral lines since hydrogen interactions are different for different spins.
   Insofar as the interaction may be described by the s-wave scattering length
   %-- reasonable, given the    diluteness of the defect ``gas" compared to microscopic length scales at sufficiently low temperatures --
   the $a$-$d$ line will be shifted by~\cite{oktel:cs-ref,oktel:cs-ref2,pethick:p-s-book,fried:hydrogen}
\be
\delta \omega &=& -\frac{4\pi \hbar^2}{m} g_2(0)\lp a_{\uparrow\downarrow}-a_{\downarrow\downarrow}\rp \expec{n}\label{eq:collision-shift}
\ee
where $g_2(\v{r}) \equiv \expec{\psi^\dagger(\v{r})\psi^\dagger(0)\psi(0)\psi(\v{r})}/\expec{n}^2$, $\expec{n}$ is the average density, and $a_{\uparrow\downarrow}$, $a_{\downarrow\downarrow}$ are the $a$-$a$ and $b$-$b$ scattering lengths, respectively.
For a noninteracting BEC  $g_2(0)=1$ while for a normal gas $g_2(0)=2$. For a partially condensed gas one in fact sees a bimodal spectrum with two peaks: one from the condensed atoms and one from the noncondensed atoms.
%Thus, a finite-temperature condensate displays a characteristic double-peaked spectrum.  This may be robust even with severe inhomogeneity.
This technique revealed BEC in magnetically \textit{trapped} spin-polarized hydrogen gas~\cite{fried:hydrogen}.
%Unfortunately the weak atomic hydrogen interactions make this spectral feature extremely difficult to observe.

\textbf{Thermodynamics and collective excitations.}
The BEC phase transition can in principle be directly observed by monitoring thermodynamic quantities such as specific heat.  Due to the small number of H atoms, the signal should be quite small.
Similarly, the presence of a superfluid component would lead one to expect a second-sound mode, which could be excited (for example) via localized heating of the sample.

\subsection{\textit{Local} Bose-Einstein condensation}

\subsubsection{Idea}

Next we pursue the idea of ``local BEC", where the defects aggregate  in small disconnected regions, each of which contains a condensate, but which have no relative phase coherence.

\subsubsection{Phenomena explainable}

This model can explain the non-Boltzmann ratio $n_b/n_a$, and the slow transport observed in the hole burning experiments.  It does not provide an explanation of the slow recombination, failure to saturate, or fast Overhauser relaxation.

\subsubsection{Consistency with experiment}
The arguments from Section~\ref{sec:Bose} about the polarization go through without change.  The slow recovery in the hole burning experiment is readily explained if the disconnected condensates are smaller than 0.1mm.
Furthermore, local clusters are naturally expected if there is the dramatic sort of concentrating mechanism described in Section~\ref{sec:Bose}.

 \subsubsection{Microscopic mechanism}

An attractive interaction between defects can lead to clustering.  A long distance phonon mediated attraction is expected for this system~\cite{mahan_2010}.  Inhomogeneities in the molecular sample, or its environment could also lead to clustering.  For example, it has been observed that the ortho and para molecules phase separate.  Furthermore, the surface that the sample sits on creates inhomogeneous strains, which couple to defects.  This can possibly cause them to accumulate.

 \subsubsection{Experimental predictions}

The temperature dependence calculated for global BEC is unchanged for local BEC,  and one again expects a double-peaked ESR spectrum.  Jointly observing these would provide a ``smoking gun" for local BEC.
We note that
at sufficiently cold temperatures the puddles
phase
lock
giving a global BEC.
However, the transition temperature would be effectively zero since atoms would have to tunnel macroscopic distances between concentrated regions.
The local concentration of H$\uparrow$, regardless of BEC, may be diagnosed by examining the dipolar shift of spectral lines due to H-H interactions.

\subsection{Nuclear spin dependent Density-of-states\label{sec:densityofstates}}
Here we outline a non-BEC scenario for the anomalous observations.

\subsubsection{Idea}

If the degeneracy of the $a$ and $b$ states were $g_a$ and $g_b$, one would expect that $n_a/n_b=(g_a/g_b)e^{\beta \Delta_{ab}}$.  Thus if a mechanism could be found to enhance $(g_a/g_b)$, then one could explain the observed ratio of $n_a/n_b$.
Assuming such a relative enhancement of the density of states, a strong RF field would lead to a saturated ratio $(n_a/n_b)_\text{sat}=g_a/g_b$.

\subsubsection{Phenomena explainable}
Both the equilibrium $n_a/n_b$ and the RF saturated $(n_a/n_b)_\text{sat}$ can be explained by this model.  We also provide a scenario whereby the Overhauser relaxation is enhanced.  Within this model, the low recombination rates could be a consequence of the defects being immobile.

\subsubsection{Consistency with experiment}

At strong excitation powers the polarization $p=(n_{\text{tot},a}-n_b)/(n_{\text{tot},a}+n_b)$ saturates to
\be
p_{\text{sat}} &=& \frac{g-1}{g+1} \label{eq:p-sat-1}
\ee
where $g\equiv g_a/g_b$.
Meanwhile, the thermal polarization is
\be
p_{\text{therm}} &=& \frac{g \exp\lp\Delta_{ab}/(k_B T)\rp-1}{g \exp\lp\Delta_{ab}/(k_B T)\rp+1}
%\nonumber\\
    \approx \frac{g-0.75}{g+0.75}\label{eq:p-therm-1}
\ee
where the last equation holds %the value
for
%the present experiments
Ahokas \textit{et al.}'s~\cite{ahokas:solid-hydrogen-anomalies-PRL} experiments, where $\Delta_{ab}=43$mK and $T=150$mK. If one takes $g=2$ one finds $p_{\text{sat}}=0.33$ and $p_{\text{therm}}=0.45$.
Ahokas \textit{et al.}~\cite{ahokas:solid-hydrogen-anomalies-PRL} experimentally find $p_{\text{sat}} = 0.2$ and $p_{\text{therm}} = 0.5$.  One should contrast this with the naive expectation of $p_{\text{sat}}=0.5$ and $p_{\text{therm}}=0.14$.

\subsubsection{Microscopic mechanism\label{eq:symmetry argument}}

There are very few mechanisms whereby the molecular hydrogen matrix can change the degeneracies of the atomic hydrogen hyperfine states.  The most plausible
%direction to look in to produce such a mechanism,
source would be to consider nuclear spin dependent interactions with o-H$_2$.  Such interactions can be produced through spin-orbit coupling in the presence of a bias magnetic field.  Hybridization of the molecular and atomic levels could in principle lead to sufficiently drastic rearrangements of the hyperfine states to affect their degeneracy.  Such a strong interaction would presumably have other spectroscopic implications, such as a severe renormalization of the $a$-$b$ splitting.  Unfortunately, the experiments observe that the $a$-$b$ splitting is changed by only 0.1\% compared to its vacuum value.

If there is significant hybridization of the atomic and molecular states, then the symmetry which forbids the $a$-$c$ transition would generically be broken. This would be a source of the fast Overhauser relaxation.

\subsubsection{Experimental predictions}

%We argued that
%``density-of-states" scenarios can account for the anomalous equilibrium and saturation polarizations.  In the mechanism presented, the symmetry breaking connected with anomalously fast Overhauser relaxation can be connected to the spectral anomalies.

One would expect that at low temperatures, $g$ should be roughly independent of temperature.   Thus the temperature dependence of the polarization in Eq.~\eqref{eq:p-therm-1} can be compared with the experiment.  For any given mechanism additional predictions are possible. As one example, in a spin-orbit mechanism hinted at above, the physics should also manifest in the populations of the o-H$_2$ states, which can be probed spectroscopically.  Finally, under further assumptions, one may be able to predict the power dependence of the polarization saturation experiments.

\section{Other observations  \label{theory:less-pred}}

We would like to point out three other possibly relevant observations.  First,
Ceperley \textit{et al.} have shown that the surface of small p-H$_2$ clusters in a vacuum  are superfluid~\cite{ceperley:superfluid-clusters}.  Similarly, Cazorla \textit{et al.} have shown that on small length scales 2D p-H$_2$ has  superfluid correlations~\cite{cazorla:2d-solid-hydrogen-sf}.  Analogous effects are predicted for defects in solid $^4$He, including grain boundaries, dislocations, and amorphous regions~\cite{pollet:grain,boninsegni:dislocation,soyler:domain}.  Ahokas \textit{ et al.}'s  crystals are of very high quality, and it is doubtful that grain boundaries and dislocations are playing an important role.  On the other hand, one could imagine that the the interface of o- and p-H$_2$ clusters could play a similar role.

Second, it seems possible that the spectroscopic anomalies are related in some way to torsional oscillator mass decoupling observed in solid hydrogen\cite{clark:hydrogen} at similar temperatures.  It is important to note that Clark  \textit{et al.}~\cite{clark:hydrogen} ruled out \textit{global} supersolidity in their experiments by comparing the response of the system in an open and blocked annulus.  \textit{Local} supersolidity, however, can also give rise to a small period drop, which would be present in both the open and blocked annuli.
%\textit{Local} supersolidity remains a possible explanation of the period observed drop.

A third observation is that
magnetic ordering transitions --- for example ferromagnetism --- would alter the ratio of $a$-state to $b$-state population. One can eliminate  ferromagnetic ordering, as it would give rise to an observable energy shift.  On the other hand, perhaps a different form of magnetic order is playing a role.

\section{Summary, conclusions, and consequences}

We reviewed the unexplained phenomena seen in experiments of Ahokas \textit{et al.}~\cite{ahokas:solid-hydrogen-anomalies-PRL}.  We enumerated a number of possible mechanisms which could be involved in producing the observed phenomena.  In particular, we gave detailed consideration to the idea, first introduced in Ref.~\cite{ahokas:solid-hydrogen-anomalies-PRL} that the non-Boltzmann ratio $n_a/n_b$ may be due to Bose-Einstein condensation of atomic hydrogen.  We conclude that global Bose-Einstein condensation is not consistent with other experimental observations.

Although we present several other scenarios, we find that none of them are wholly satisfactory.  Although some of the phenomena can be explained by local BEC, it fails to provide a mechanism for the unexpected saturation population $(n_a/n_b)_{\rm sat}$ when a strong RF field is applied.  We do find that all of the phenomena would be consistent with a nuclear spin dependent density of states.  However, we are unable to provide a microscopic mechanism for this density of states.

Ultimately, substantial experimental work will be necessary to clarify the situation.  Our arguments make a strong case that measuring the polarization's temperature dependence is a promising first step, and suggests other experimental signatures  --- especially in transport and spectral features --- that would clarify the phenomena.

During the preparation of this paper, new results came out from Ahokas \textit{et al.}~\cite{newahokas}, which introduced new mysteries.  In particular they observe substantial density and substrate dependence of the population ratio $n_a/n_b$.  All of our considerations remain valid, with the additional clue that whatever the underlying mechanism is, it must involve the surface of the sample, and be sensitive to density.  For example, the formation of superfluid domains could be influenced by the substrate, or magnetic impurities in the substrate could interact with atomic hydrogen, leading in some way to the unexpected density of states.

\textbf{Acknowledgments.}  We would like to acknowledge discussions with David Lee, J\"arno Jarvinen, Sergei Vasiliev, and Cyrus Umrigar.  This work was supported by NSF Grant PHY-0758104.

%\bibliographystyle{prsty}
%\bibliography{atomic-H2-in-crystals}{}

\end{document}